\documentclass[fleqn,usenatbib]{mnras}

\usepackage{newtxtext,newtxmath}

\usepackage[T1]{fontenc}

\DeclareRobustCommand{\VAN}[3]{#2}
\let\VANthebibliography\thebibliography
\def\thebibliography{\DeclareRobustCommand{\VAN}[3]{##3}\VANthebibliography}

\usepackage{graphicx}	
\usepackage{amsmath}	

\defcitealias{Lin:2021a}{L21}

\title[Attenuation properties at $z\sim 7$]{Effects of dust sources on dust attenuation properties in IllustrisTNG galaxies at $z\sim 7$}

\author[Y.-M. Hsu et al.]{
Yuan-Ming Hsu,$^{1,2}$\thanks{E-mail: technic960183@gmail.com}
Hiroyuki Hirashita,$^{2,3}\thanks{E-mail: hirashita@asiaa.sinica.edu.tw}$
Yen-Hsing Lin,$^{4}$
Peter Camps$^{5}$
and Maarten Baes$^{5}$
\\
$^{1}$Department of Physics, National Taiwan University, No. 1, Section 4, Roosevelt Road, Taipei 10617, Taiwan\\
$^{2}$Institute of Astronomy and Astrophysics, Academia Sinica, Astronomy-Mathematics Building, No. 1, Section 4, Roosevelt Road, Taipei 10617, Taiwan\\
$^{3}$Theoretical Astrophysics, Department of Earth and Space Science, Osaka University, 1-1 Machikaneyama, Toyonaka, Osaka 560-0043, Japan\\
$^{4}$Institute of Astronomy, National Tsing Hua University, 101, Section 2, Kuang-Fu Road, Hsinchu 30014, Taiwan\\
$^{5}$Sterrenkundig Observatorium, Universiteit Gent, Krijgslaan 281 S9, B-9000 Gent, Belgium
}

\date{Accepted XXX. Received YYY; in original form ZZZ}

\pubyear{2022}

\begin{document}
\label{firstpage}
\pagerange{\pageref{firstpage}--\pageref{lastpage}}
\maketitle

\begin{abstract}
Dust emission from high-redshift galaxies gives us a clue to the origin and evolution of dust in the early Universe. Previous studies have shown that different sources of dust (stellar dust production and dust growth in dense clouds) predict different ultraviolet (UV) extinction curves for galaxies at $z\sim 7$ but that the observed attenuation curves depend strongly on the geometry of dust and star distributions. Thus, we perform radiative transfer calculations under the dust--stars geometries computed by a cosmological hydrodynamic simulation (IllustrisTNG). This serves to investigate the dust attenuation curves predicted from `realistic' geometries. We choose objects with stellar mass and star formation rate appropriate for Lyman break galaxies at $z\sim 7$. We find that the attenuation curves are very different from the original extinction curves in most of the galaxies. This makes it difficult to constrain the dominant dust sources from the observed attenuation curves. We further include infrared dust emission in the analysis and plot the infrared excess (IRX)--UV spectral slope ($\beta$) diagram. We find that different sources of dust cause different IRX--$\beta$ relations for the simulated galaxies. In particular, if dust growth is the main source of dust, a variation of dust-to-metal ratio causes a more extended sequence with smaller IRX in the IRX--$\beta$ diagram. Thus, the comprehensive analysis of the abundances of dust and metals, the UV slope, and the dust emission could provide a clue to the dominant dust sources in the Universe.
\end{abstract}




\section{Introduction}
Recent observations by the Atacama Large Millimetre/submillimetre Array (ALMA) have detected dust emission from galaxies, represented by Lyman break galaxies (LBGs), in the early Universe, and the redshift frontier of dust emission observations has expanded to $z\gtrsim 7$, where $z$ is the redshift \citep[e.g.][]{Hodge_DaCunha:2020, Schouws:2022a}. These observations in principle can provide clues to the first dust enrichment or the origin of dust in the Universe. Since dust is an important component in the interstellar medium (ISM) in terms of radiative and chemical processes \citep[e.g.][]{Draine:2009a}, revealing the origin of dust has a large impact on the understanding of the evolution of the ISM in galaxies.

One of the important roles of dust in galaxies appears in radiative processes. Dust absorbs and scatters stellar light in the ultraviolet (UV)--optical and reprocesses it in the infrared (IR). Thus, dust modifies the shape of the spectral energy distributions (SEDs) of galaxies. The sum of absorption and scattering is referred to as the extinction, and the wavelength dependence of extinction, called extinction curve, is of fundamental importance in regulating the UV--optical SED shape \citep[e.g.][]{da-Cunha:2008a,Boquien:2019a,Abdurrouf:2021a}. The grain size distribution as well as the dust composition is important in determining the extinction curve \citep[e.g.][]{Mathis:1977a}. This means that we need to understand not only the total amount of dust but also the grain size distribution.

There are some predictions on the grain size distribution for LBGs at $z\gtrsim 7$. \citet{Liu:2019a} used data of ALMA-detected LBGs at $z\gtrsim 7$ in the literature to constrain the dust enrichment history, and found that efficient dust condensation in stellar ejecta and fast dust growth by the accretion of gas-phase metals both explain the observed dust masses equally well \citep[see also][]{Wang:2017a}. In particular, non-stellar dust sources (accretion) is also proposed to explain the dust masses of some high-redshift galaxies \citep{Mancini:2015a,Lesniewska:2019a}. \citet{Liu:2019a} also calculated the grain size distributions, which proved to be significantly different between the two dust sources (stellar dust production vs.\ dust growth by accretion). If the dust abundance is dominated by stellar dust production (mainly supernovae at high redshift), the grain sizes are biased to large (sub-micron) radii \citep[see also][]{Nozawa:2007a,Gall:2014a}. In contrast, if the dust mass increase is predominantly caused by accretion, small ($\lesssim 0.01~\micron$) grains are dominant because of their high accretion efficiencies (i.e.\ large surface-to-volume ratios; see also \citealt{Weingartner:1999a,Hirashita:2011a}). In the first scenario where the dust is dominated by stellar production (stardust scenario), optical--UV extinction curves are flat because of large grain sizes. In contrast, the second scenario with dust growth by accretion (dust growth scenario) predicts steep extinction curves. This means that extinction curves may provide a useful guide to distinguish the dominant dust sources in high-redshift galaxies.

As long as we observe the integrated emission from a galaxy, the stellar light is not simply attenuated following the extinction curve, but is affected by various radiative transfer effects within the galaxy \citep[e.g.][]{Calzetti:2001a,Salim:2020a}. The wavelength dependence of the attenuation is referred to as the attenuation curve. Therefore, if we analyze the SED of a galaxy, what we can extract in terms of the dust property is not the extinction curve but the attenuation curve \citep[e.g.][]{Buat:2011a,Kriek:2013a}. It is well known that the attenuation curve is strongly modified by the spatial distribution geometries of dust and stars \citep[dust--stars geometries; e.g.][]{Witt:1996a,Witt:2000a,Granato:2000a}.

\citet[][hereafter L21]{Lin:2021a} examined the spherically symmetric geometries for dust and star distributions and calculated the attenuation curves for the above-mentioned two different grain size distributions predicted from different grain sources -- stardust scenario and dust growth scenario. They found that attenuation curves tend to be flat if dust and stars are distributed in a well mixed way \citep[see also][]{Narayanan:2018a}. \citetalias{Lin:2021a} called this flattening effect `\textit{geometry effect}' following \citet{Narayanan:2018a}. This geometry effect is also found and analyzed by \citet{DiMascia:2021a}. In contrast, attenuation curves tend to be steepened if young stars are more embedded than old stars, which \citetalias{Lin:2021a} called `\textit{age effect}' \citep[see also][]{Charlot:2000a,Narayanan:2018a}, or if scattering is important (`\textit{scattering effect}'; \citealt{Witt:2000a,Baes:2001a,Goobar:2008a}). \citetalias{Lin:2021a} showed that these complicated effects strongly modify the shape of attenuation curve, and concluded that it is difficult to discriminate the grain sources only using the attenuation curves.

The extinction properties could be constrained further if we include IR emission into the analysis. The relation of two attenuation indicators, IR excess (IRX) and UV SED slope ($\beta$), is often used to analyze the extinction (or attenuation) properties of galaxies \citep{Meurer:1999a}. This IRX--$\beta$ relation is particularly useful for high-redshift galaxies \citep[e.g.][]{Casey:2014a} because it is not easy to extract the attenuation curves from a limited number of observational photometric points. Indeed, many studies have plotted IRX--$\beta$ relations for $z\gtrsim 7$ LBGs \citep[e.g.][]{Hashimoto:2019a,Schouws:2022a}. Theoretical studies by \citet{Popping:2017a} and \citet{Narayanan:2018b} showed that the above radiative transfer effects in addition to the original extinction curve shape affect the IRX--$\beta$ relation. Although the uncertainties are large, LBGs at $z>5$ seem to be dispersed in the IRX--$\beta$ diagram \citep[see also][]{Capak:2015a,Faisst:2017a,Fudamoto:2020a} unlike nearby galaxies which lie more or less on a sequence \citep[e.g.][]{Meurer:1999a,Takeuchi:2012a}. \citetalias{Lin:2021a}, based on the above radiative transfer calculations, examined the IRX--$\beta$ relation and found that the two different dust sources (stardust and dust growth scenarios) predict different positions in the relation. The dust growth scenario, which has a steeper extinction curve than the stardust scenario, predicts systematically larger values of $\beta$ (i.e.\ redder UV spectra) at a fixed value of IRX. This is because with a fixed amount of dust attenuation, the steeper extinction curve extinguishes radiation at short wavelengths more efficiently.

It is still necessary to investigate the effect of dust distribution geometry using more realistic structures. \citetalias{Lin:2021a} assumed spherical symmetry in order to facilitate the physical interpretation. The spatial distribution of dust is driven by hydrodynamics because dust is coupled well with gas dynamics on (sub)galactic scales \citep{McKinnon:2018a}. Thus, hydrodynamic simulations provide us with useful tools to investigate the dust effects in realistic dust distribution geometries. In addition, statistical properties of dust attenuation is important to obtain a general conclusion not strongly influenced by specific geometries realized in a small number of galaxies. Cosmological hydrodynamic simulations, which compute the evolution of cosmic structures including dark matter and baryons, are in general able to predict statistical properties of galaxies in the Universe \citep[e.g.][]{Springel:2003a}. These simulations also include detailed baryonic physics such as star formation and energy input (feedback) from stars and super-massive black holes, enabling us to predict evolution of various galactic properties (star formation history, metal enrichment, multiphase ISM structures, etc.) on (sub-)kiloparsec scales \citep[e.g.][]{Pillepich:2018a}. In this paper, we use a galaxy sample in a cosmological hydrodynamic simulation to investigate dust attenuation in predicted spatial distributions of dust and stars. Specifically, we use The Next Generation Illustris project (hereafter IllustrisTNG or TNG, see Section \ref{subsec:TNG} for more information). The importance of considering realistic structures of dust distribution is further pronounced by the actually observed inhomogeneous dust--star distributions in high-redshift galaxies \citep{Bowler:2022a,Inami:2022a}.

There have been some studies of dust attenuation properties in high-$z$ galaxies (including the redshift of interest, $z\sim 7$) using cosmological simulations. \citet{Yajima:2014a} performed a cosmological zoom-in simulation and showed that the difference in dust grain sizes has an imprint on the observed UV luminosity function through different wavelength dependence of dust extinction. \citet{Ma:2019a} also used a cosmological zoom-in simulation, and found a clear IRX--$\beta$ relation predominantly determined by the adopted extinction curve. They concluded that the IRX--$\beta$ relation is more strongly determined by the shape of the adopted extinction curve than by the dust-to-metal ratio (or the normalization of the attenuation law). \citet{Liang:2021a} found similar IRX--$\beta$ relations at $z\sim 6$, using a cosmological zoom-in simulation \citep[see also][]{Liang:2019a}. They also pointed out that the intrinsic stellar spectrum, or the stellar age is important in determining the position (or redshift evolution) in the IRX--$\beta$ relation \citep[see also][]{Burgarella:2020a}. \citet{Vijayan:2022a} also performed a cosmological zoom-in simulation, and found an IRX--$\beta$ relation similar to that of nearby starburst galaxies (implying a Calzetti-like attenuation curve), although they assumed the Small Magellanic Cloud (SMC) extinction curve. \citet{Pallottini:2022a}, using their on-the-fly radiative transfer code, calculated the radiative properties of galaxies using a cosmological zoom-in simulation. They found a large IRX with relatively blue UV colours (i.e.\ small $\beta$). They suggested that a spatial segregation between UV- and IR-dominated regions is responsible for the position of their simulated galaxies in the IRX--$\beta$ diagram \citep[see also][]{Behrens:2018a}. The above two studies indicate that the dust--stars geometry play an important role in shaping the attenuation curve (or the IRX--$\beta$ relation) and that the attenuation curve does not necessarily reflect the original extinction curve. \citet{Shen:2022a} used IllustrisTNG and focused on SEDs, luminosity functions, dust temperatures, but did not analyze the dust attenuation properties in detail. All the above studies basically fixed the dust property (or the extinction curve), and it is not clear how different dust properties (or extinction curves), which are expected for high-redshift galaxies \citep{Liu:2019a}, affect the dust attenuation curves and the IRX--$\beta$ relation.

In this paper, we focus on the effect of different extinction properties on dust attenuation curves and IRX--$\beta$ relations, motivated by the above mentioned results that different dust sources (stardust scenario vs.\ dust growth scenario) predict different extinction curves. Although \citetalias{Lin:2021a} already investigated this effect in idealized spherical geometries, we extend their studies by using `realistic' dust--star geometries predicted from a cosmological simulation. In this way, we are able to clarify if different dust sources have any significant imprint on the observed attenuation properties in the spatial distributions of dust and stars realized as a result of the cosmological structure formation. At the same time, we also examine if observed IRX--$\beta$ relations for galaxies at $z\sim 7$ (the frontier redshift for dust observation) are explained by the grain size distributions expected from dust evolution models.

This paper is organized as follows. In Section \ref{sec:model}, we explain the cosmological simulation, the dust models, the post-processing radiative transfer calculations, and the outputs that quantify the attenuation properties of simulated galaxies. In Section \ref{sec:result}, we present the results. In Section \ref{sec:discussion}, we discuss uncertainties and implications of the results. In Section \ref{sec:conclusion}, we summarize our findings. We follow the same cosmological model as used in TNG ($h = 0.6774$, $\Omega_\Lambda = 0.6911$, $\Omega_\mathrm{m}= 0.3089$, and $\Omega_\mathrm{b} = 0.0486$; \citealt{Planck:2016a}).

\section{Methods}\label{sec:model}

\subsection{IllustrisTNG and sample selection}\label{subsec:TNG}

We model the spatial distributions of dust and stars in galaxies at $z\sim 7$ using the result of a state-of-the-art cosmological hydrodynamic simulation in the IllustrisTNG project \citep{Nelson:2019a}. There are a variety of simulations with different cubic volumes with comoving side lengths of $\sim 50$, 100, and 300 Mpc, referred to as TNG50 \citep{2019MNRAS.490.3234N,2019MNRAS.490.3196P}, TNG100, and TNG300 \citep{Marinacci:2018a,Naiman:2018a,2018MNRAS.475..624N,Pillepich:2018a,Springel:2018a}, respectively. Each of these simulations solves for the gravitational evolution of dark matter and baryons from $z=127$ to the present day, including baryonic physics and energy feedback from formed structures \citep{Weinberger:2017a,Pillepich:2018b}. 

\begin{figure}
    \centering
    \includegraphics[width=\linewidth]{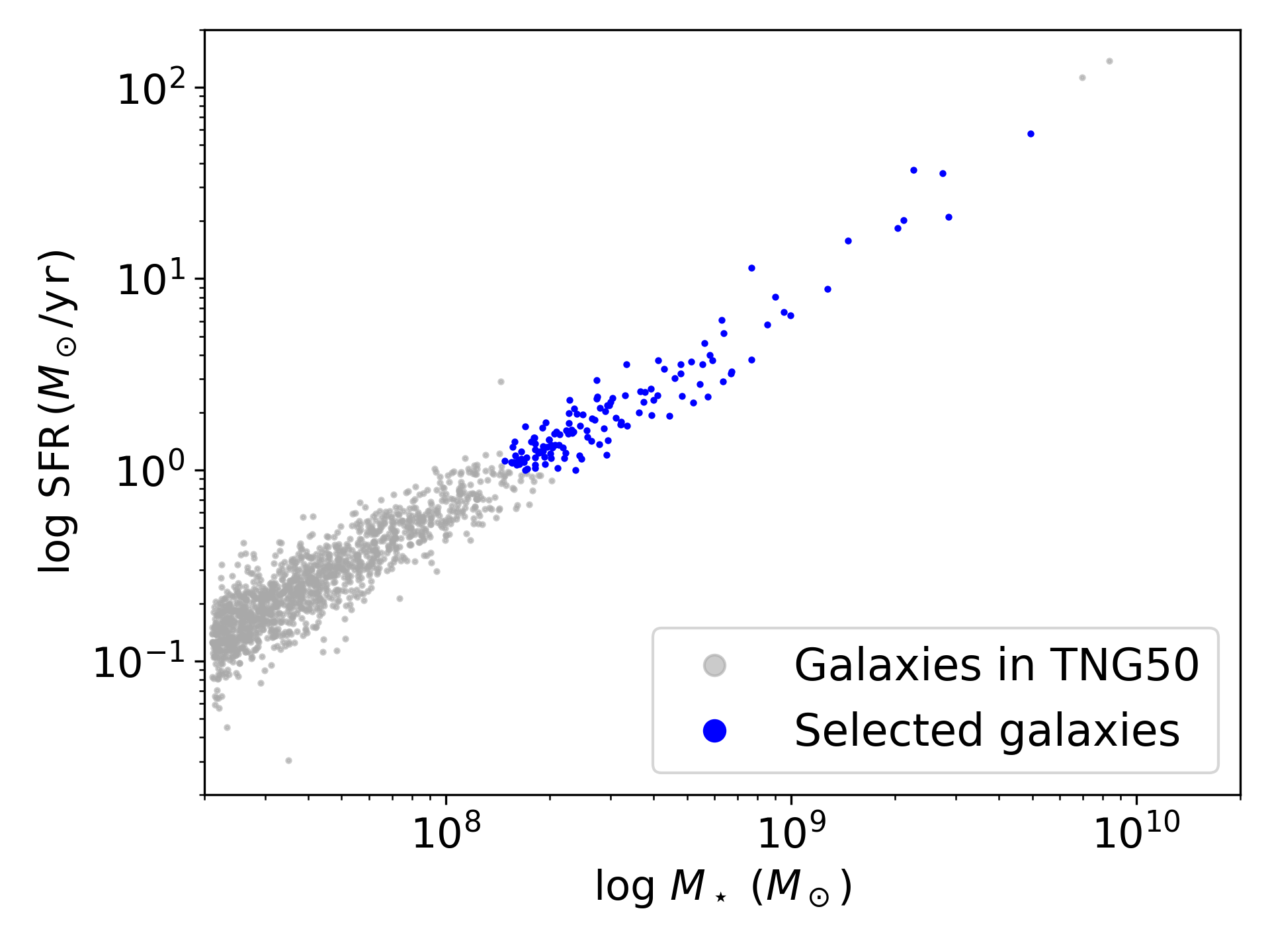}
    \caption{Stellar mass ($M_\star$)--SFR relation of the galaxies in the $z=7.0$ snapshot in TNG50. The selected objects in this paper are shown in blue.}
    \label{fig:sample_dist}
\end{figure}

Since the radiative transfer results could be sensitive to the spatial resolution, we adopt the highest resolution simulation, TNG50-1 (hereafter TNG50). High resolution is also useful to select low-mass galaxies. The mass resolutions of baryons and dark matter are $5.74\times10^4h^{-1}~\mathrm{M}_{\sun}$ and $3.07\times10^5h^{-1}~\mathrm{M}_{\sun}$, respectively. As mentioned in the Introduction, we focus on LBGs at $z\gtrsim 7$, where ALMA detected the most distant dust continuum. Thus, we adopt the snapshot at $z=7.0$ in TNG50. For high-redshift galaxies at $z\sim 7$ whose dust emission has detected by ALMA, as compiled by \citet{Liu:2019a}, the range of the stellar mass ($M_\star$) is $10^9$--$10^{10}$ M$_{\sun}$, and that of the star formation rate (SFR) is 10--100 M$_{\sun}$ yr$^{-1}$. We expand these parameter ranges downwards by an order of magnitude to include non-detected galaxies and finally choose galaxies with stellar mass of $1.5\times10^8$--$1.5\times10^{10}$ M$_{\sun}$\footnote{The range also takes into account the available computational resources. We expect that further lowering the lower mass limit to e.g.\ $1.0\times 10^8$ M$_{\sun}$ has little impact on the discussions and conclusions in this paper.} and a SFR of 1--100 M$_{\sun}$ yr$^{-1}$. This expansion is useful to examine the systematic variation caused by the stellar mass (given that $M_\star$ and SFR are broadly correlated; Fig.\ \ref{fig:sample_dist}). The lowest mass ($M_\star =1.5\times10^8$ M$_{\sun}$) still contains a large number ($\gtrsim 10^4$) of baryonic particles, which helps to obtain a meaningful result for radiative transfer calculations. With the above criteria, we select 133 galaxies from the $z=7$ snapshot. We show the relation between SFR and $M_\star$ of the selected galaxies in Fig.\ \ref{fig:sample_dist}. We observe that we almost completely include galaxies with $M_\star >1.5\times 10^8$ M$_{\sun}$ because of the strong correlation between $M_\star$ and SFR; in other words, the criterion on SFR does not significantly affect the sample properties.

We tested the resolution effect by using TNG100: We chose 52 galaxies with the following criteria: $M_\star=1.5\times10^9$--$1.5\times10^{10}$ M$_\odot$ and SFR = 10--100 M$_\odot$ yr$^{-1}$. The lower bounds of these criteria for TNG100 is set to 10 times larger than those in TNG50 because of worse mass resolution. We confirmed that the statistical distributions of attenuation curves and IRX--$\beta$ relations are similar between TNG50 and TNG100 without any significant systematic deviations in the common stellar mass range. Since the extension toward lower stellar mass achieved by TNG50 (as well we higher mass resolution) is useful, we only discuss the results for TNG50 in this paper.

\subsection{Dust model}\label{subsec:dust}

We adopt the same dust model as in \citetalias{Lin:2021a}. The input extinction curves are based on the grain size distributions predicted by \citet{Liu:2019a} for $z\gtrsim 7$ LBGs. They used the evolution model of grain size distribution developed by \citet{Hirashita:2019a} \citep[originally from][]{Asano:2013a}, who included stellar dust production, dust destruction by supernovae, grain growth by accretion and coagulation, and grain disruption by shattering in a consistent manner with the metal enrichment by stars. \citet{Liu:2019a} predicted the resulting grain size distributions in the following two scenarios: efficient stellar dust production (hereafter referred to as \textit{the stardust scenario}) and fast dust growth in the ISM (\textit{the dust growth scenario}), and showed that both scenarios equally reproduce the stellar and dust masses and the SFRs actually observed for $z\gtrsim 7$ LGBs at a typical age of $\sim 10^8$ yr. However, the grain size distributions are significantly different. In the stardust scenario, the grain size distribution is dominated by large ($a\gtrsim 0.1~\micron$) grains, which reflect the typical sizes of grains formed by stellar sources \citep[][and references therein]{Liu:2019a}. In the dust growth scenario, in contrast, small ($a\lesssim 0.01~\micron$) grains are dominant since their larger surface-to-volume ratios lead to more efficient accretion of the gas-phase metals. Note that \citetalias{Lin:2021a} recalculated the grain size distributions using the \citet{Chabrier:2003a} initial mass function (IMF) instead of the \citet{Salpeter:1955a} IMF. This change does not cause any significant difference in our results.

To calculate the extinction curve, the grain size distribution in each scenario is decomposed into two dust species, silicate and graphite following \citetalias{Lin:2021a}: To focus on the difference between the two scenarios, we fix the dust composition by assuming a constant value of 0.9 for the mass ratio of silicate to the total dust mass ($f_\mathrm{sil}$). This value is appropriate at typical young ages of the high-redshift LBGs \citep{Hirashita:2020a,Huang:2021a}. We display the effect of $f_\mathrm{sil}$ in Appendix \ref{apdx:sgratio}, and also mention it in Section \ref{subsec:dm}. The resulting extinction curves are very different in the two scenarios with the dust growth scenario predicting a much steeper curve than the stardust scenario. The extinction normalized to the value at $\lambda =0.3~\micron$ ($\lambda$ is the rest-frame wavelength) is four times larger in the dust growth scenario than in the stardust scenario at $\lambda =0.1~\micron$ (see \citetalias{Lin:2021a} for details).

We determine the mass of dust ($m_\mathrm{dust}$) for each discrete gas particle in the simulation using the metallicity and the derived temperature provided by TNG as
\begin{equation}
    m_\mathrm{dust}=
    \begin{cases}
        R_\mathrm{D/M}m_\mathrm{gas}Z_\mathrm{gas} & \text{if $T_\mathrm{gas}<T_\mathrm{th}$,} \\
        0 & \text{if $T_\mathrm{gas}\ge T_\mathrm{th}$},
    \end{cases}
\end{equation}
where $T_\mathrm{gas}$ is the gas temperature, $m_\mathrm{gas}$ is the gas mass, $Z_\mathrm{gas}$ is the metallicity of the gas, $R_{\text{D/M}}$ is the dust-to-metal ratio, and $T_\mathrm{th}$ is the threshold temperature above which dust is destroyed. TNG provides $m_\mathrm{gas}$ and $Z_\mathrm{gas}$, while we give $R_\mathrm{D/M}$ as a free parameter. We adopt a threshold of $T_\mathrm{th}=10^5$ K which is well above the typical temperature of the diffuse ISM ($\sim 10^4$ K). The threshold mimics the dust destruction in the circum-galactic medium \citep{Aoyama:2018a} and in the ISM directly heated by supernovae \citep{Hu:2019a}. The above threshold is also near to the value ($7.5\times10^4$ K) adopted by \citet{schulz2020redshift}. In nearby galaxies such as the Milky Way, the dust-to-metal ratio is $\sim 0.3$ \citep[e.g.][]{DeVis:2019a,Chiang:2021a}, and the ALMA-detected LBGs at $z\sim 7$ have similar dust-to-metal ratios \citep{Liu:2019a,Dayal:2022a}. However, this value is uncertain. Some studies assume dust-to-metal ratio of $\sim 0.1$ for the dust condensation efficiency in stellar ejecta based on theoretical results \citep[e.g.][]{Inoue:2011a,Kuo:2013a}. A similar dust-to-metal ratio (0.08) is preferred by \citet{Behrens:2018a}, who compared their radiative transfer calculations with a galaxy at $z=8.38$. Recent studies suggested that the dust destruction by supernovae could be efficient in diminishing the dust content in high-redshift galaxies \citep{Burgarella:2020a,Nanni:2020a}. In this case, the dust-to-metal ratio is expected to be lower. Therefore, to account for the above ranges and uncertainties of dust-to-metal ratios, we examine $R_\mathrm{D/M}=0.3$, 0.1, and 0.05, but still assume $R_\mathrm{D/M}$ to be constant. The grain size distribution is normalized so that the total dust mass becomes $m_\mathrm{dust}$ on each gas particle.

Some studies of nearby galaxies \citep[e.g.][]{RemyRuyer2014gas,DeVis:2019a,Galliano:2021a} indicate that the dust-to-metal ratio depends on the metallicity. A metallicity-dependent $R_\mathrm{D/M}$ tends to enhance the variation of dust-to-gas ratio; thus, it tends to expand the variation in the attenuation curves and the IRX--$\beta$ relation. However, we confirmed that, because of the large variety in the gas distribution geometries among the sample galaxies, the metallicity dependence of $R_\mathrm{D/M}$ only has a subdominant effect on the scatters in the attenuation curves and the IRX--$\beta$ relation. The above change of the constant $R_\mathrm{D/M}$ values (0.05--0.3) has a larger influence on these properties. Moreover, the dust attenuation is more prominent in regions with high metallicity, where the constant $R_\mathrm{D/M}$ is justified \citet{RemyRuyer2014gas}. Therefore, we assume $R_\mathrm{D/M}$ to be constant but still examine the effect of its variation.

\subsection{Radiative transfer simulation}\label{subsec:RT}

For the radiative transfer calculation, we use \textsc{skirt9}, which is based on the Monte Carlo method \citep{CAMPS2020100381}. In addition to the stellar radiation transfer in a dusty medium, it calculates the secondary emission from dust in a consistent manner with the radiative energy absorbed by the dust, including iterations for dust self-absorption. We import the gas and star particles of each galaxy (Section \ref{subsec:TNG}) into the \textsc{skirt9} module. Both types of particles contain the information of position, smoothing length, mass and metallicity. The stellar age is also available for each stellar particle. These quantities are used to obtain the necessary inputs for the radiative transfer calculations as described below. The grain size distribution (including the appropriate normalization for the total dust abundance determined by $R_\mathrm{D/M}$; Section \ref{subsec:dust}) in each gas particle is included into \textsc{skirt9} by the newly implemented grain size distribution feature.

For the numerical implementation of radiative transfer, \textsc{skirt9} uses adaptive mesh refinement (AMR) grids to store the radiation field at each spatial location. If a grid contains more than $10^{-6}$ of the total mass of the dust in the whole simulation box, we adopt a higher resolution level of the AMR. The highest spatial resolution is $\sim 2$--4~pc, comparable to or smaller than the spatial resolution achieved by TNG50. In this work, there are roughly a few millions of AMR girds in total for each galaxy.

For the instinsic stellar spectrum of each stellar particle, the \citet{Bruzual:2003a} model is adopted based on the metallicity and stellar age given by TNG for stellar particles with age $>10$~Myr. We adopt the \citet{Chabrier:2003a} IMF. Since we are only interested in the far-IR wavelengths where dust in radiative equilibrium dominates the emission, we do not consider stochastic heating. We adopt the optical properties of silicate and graphite from the \citet{Draine:1984a} model and the calorimetric properties from the \citet{Draine:2001a} model. For the stellar particles of which the age is less than $10\ \text{Myr}$, because they are usually associated with dense star-forming regions that TNG is not able to resolve, we adopt the MAPPINGS III model \citep{Groves:2008a}. This prescription for young stars was also adopted by \citet{Camps:2018a}, and the time-scale (10 Myr) of this `embedded' phase is consistent with that derived from the analysis of the UV properties in high-redshift LBGs by \citet{Mancini:2016a}. For the MAPPINGS III SED, several parameters need to be fixed. The metallicity is given by TNG and the SFR is simply assumed to be constant over the past $10~\text{Myr}$. The compactness of the star clusters $C$, the gas pressure, and the photodissociation region covering factor are set to $10^5$, $1.38\times10^{-12}~\text{Pa}$, and 0.1, respectively \citep{Groves:2008a,Jonsson:2010a}.

Note that the dust extinction law implicitly assumed in the MAPPINGS III model is not consistent with our assumption. Thus, we examined a case in which the MAPPINGS III model is not applied at all. As expected, the IRX--$\beta$ relation shifts towards lower extinction, but the shift is along the sequence; that is, the MAPPINGS III implementation does not alter the IRX--$\beta$ sequence, which is well identified in the results (Section \ref{sec:result}). The statistics of the attenuation curve shapes is not changed significantly by the inclusion of MAPPINGS III. Overall, the difference between the calculations with and without MAPPINGS III has the same effect as changing $R_\mathrm{D/M}$, whose effect is discussed in detail in Section \ref{subsec:dm}. Because of the degeneracy with $R_\mathrm{D/M}$, the inclusion of MAPPINGS III does not cause any direct or systematic influence on the conclusions drawn in this paper, as long as $R_\mathrm{D/M}$ is moved in a sufficiently large range.

As a result of the \textsc{skirt9} calculation, we obtain the radiation fluxes from the galaxy received by 14 virtual instruments placed in different directions, including both intrinsic and attenuated fluxes. The wavelengths are sampled with 200 logarithmic grid points from 0.09 to $10^4~\micron$.

\subsection{Output quantities}

As mentioned in the Introduction, our main focus is on dust attenuation; in particular, we aim at clarifying how the resulting attenuation curves are affected by input extinction curves. Observationally, it is rather easier (especially for high-redshift galaxies) to obtain the IRX--$\beta$ relation, which can be constructed only from sparse photometric data points. Following the approach taken by \citetalias{Lin:2021a}, we output both attenuation curves and IRX--$\beta$ relations. In the following subsections, we explain how to calculate these output quantities.

\subsubsection{Attenuation curves}
The attenuation at wavelength $\lambda$, denoted as $A_\lambda$, in units of magnitude is related to effective optical depth $\tau_\lambda^\text{eff}$ by the relation $A_\lambda=1.086\tau_\lambda^\text{eff}$. The effective optical depth is defined as (e.g.\ \citetalias{Lin:2021a})
\begin{equation}
\tau_\lambda^\text{eff}=\ln{\frac{L_{\lambda,\text{in}}}{L_{\lambda,\text{att}}}},
\end{equation}
where $L_{\lambda,\text{in}}$ and $L_{\lambda,\text{att}}$ represent the intrinsic and attenuated luminosities (calculated by the virtually detected flux assuming isotropic radiation) per wavelength, respectively.

\subsubsection{IRX--$\beta$ relation}
We explain how to calculate $\beta$ and IRX. We evaluate the UV slope, $\beta$, as
\begin{equation}
    \beta=\frac{\log_{10}(L_{\nu_2}/L_{\nu_1})}{\log_{10}(\lambda_2/\lambda_1)}-2,
\end{equation}
where all the quantities are evaluated in the restframe: we adopt $\lambda_1=0.16~\micron$ and $\lambda_2=0.25~\micron$ (\citetalias{Lin:2021a}) to avoid being influenced by the carbon bump at $\lambda=0.22~\micron$ (which causes an additional uncertainty in the modeling; \citealt{Safarzadeh:2017a}), and denote the luminosity density (per frequency) at these two wavelengths as $L_{\nu_1}$ and $L_{\nu_2}$ ($\nu_1$ and $\nu_2$ are the frequencies of light corresponding to $\lambda_1$ and $\lambda_2$, respectively). The IRX, which is defined on logarithmic scales in this paper, is evaluated as
\begin{equation}
    \text{IRX}=\log_{10}\left(\frac{L_\text{IR}}{L_\text{UV}}\right),
\end{equation}
where $L_\text{IR}$ is the IR luminosity integrated from 3 to $1000~\micron$, and $L_\text{UV}$ is the UV luminosity estimated by $L_\text{UV}=\nu_1L_{\nu_1}$.

\section{Results}\label{sec:result}

In this section, we show the basic results of the attenuation curves and the IRX--$\beta$ relation from the selected galaxies in TNG50. We investigate the difference in the grain size distribution (the \textit{stardust scenario} and \textit{dust growth scenario}; Section \ref{subsec:dust}), and the variation of the dust abundance regulated by the dust-to-metal ratio $R_\mathrm{D/M}$. We also examine the dependence on the stellar mass to clarify the effect of the stellar mass selection.

\subsection{Different dust sources (two scenarios)}\label{subsec:scenarios}
\begin{figure}
    \centering
    \includegraphics[width=\linewidth]{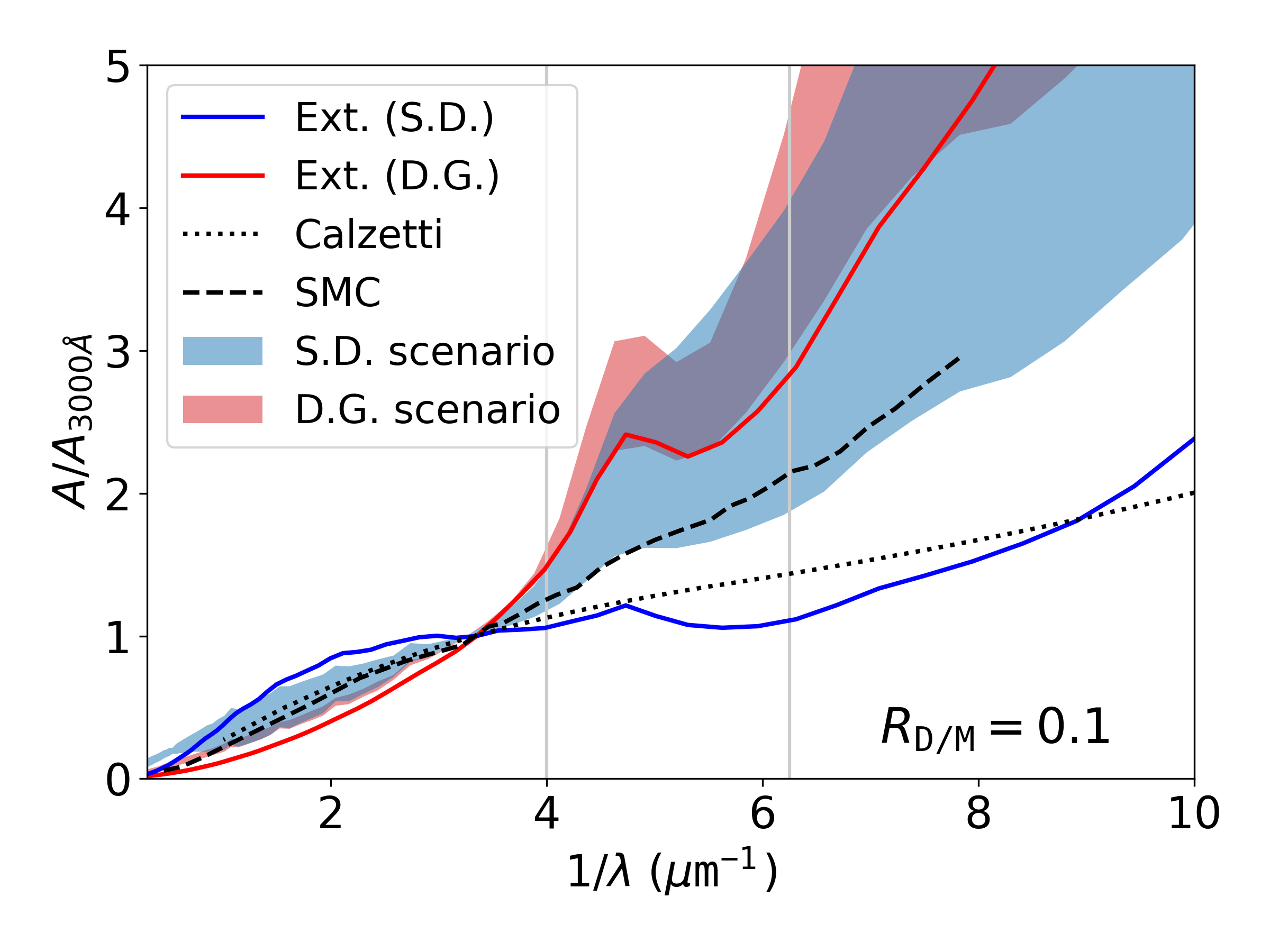}
    \includegraphics[width=\linewidth]{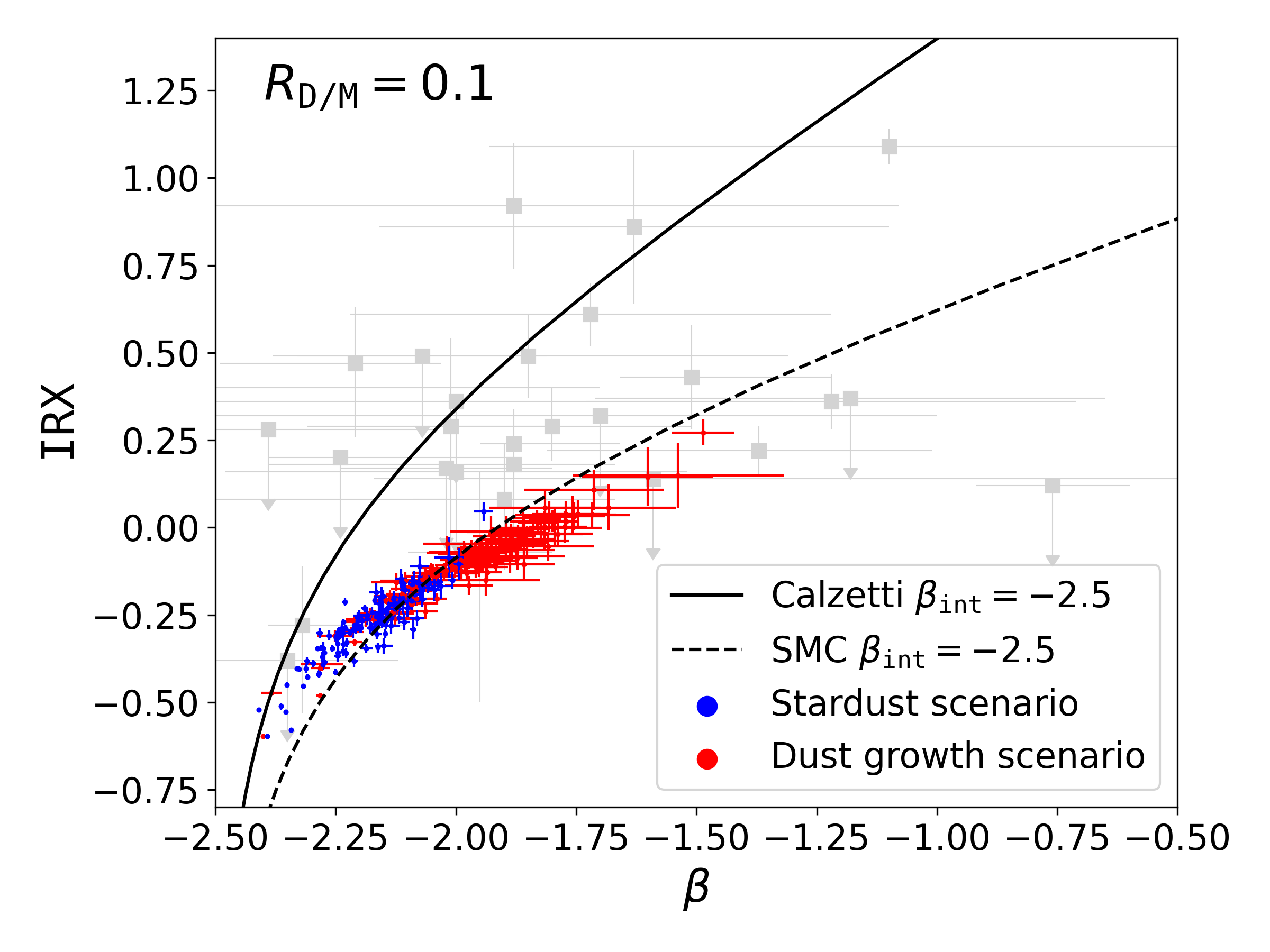}
    \caption{Upper: Attenuation curves of the $z=7$ sample in TNG with $R_\text{D/M}=0.1$, normalized to the value at rest-frame wavelength $\lambda =3000$~\AA. The blue and red areas represent the 16th--84th percentile range of the attenuation curves in the stardust and dust growth scenarios, respectively. The blue and red solid curves show the original extinction curves for the stardust and dust growth scenarios, respectively. For reference, the Calzetti attenuation curve \citep{Calzetti:2000a} and the SMC extinction curve \citep{Pei:1992a} are also shown.
    Lower: IRX--$\beta$ relations of the same galaxies in TNG50 with $R_\text{D/M}=0.1$. The blue and red points show the results for the stardust and dust growth scenarios, respectively. Each of these points represents a galaxy with error bars showing the standard deviation for the 14 different directions (inclination effect). We show the IRX--$\beta$ relations expected for the Calzetti and SMC curves with $\beta_\mathrm{int} =-2.5$ by the solid and dashed lines, respectively. The gray points are the observational data of galaxies at $z\gtrsim 7$ \citep{Hashimoto:2019a,Schouws:2022a,Bowler:2022a}.}
    \label{fig:Att_Cur}
\end{figure}

In Fig.\ \ref{fig:Att_Cur}, we focus on the difference between the two scenarios (different dust sources) for the case with the intermediate dust-to-metal ratio $R_\mathrm{D/M}=0.1$.

\begin{figure}
    \centering
    \includegraphics[width=\linewidth]{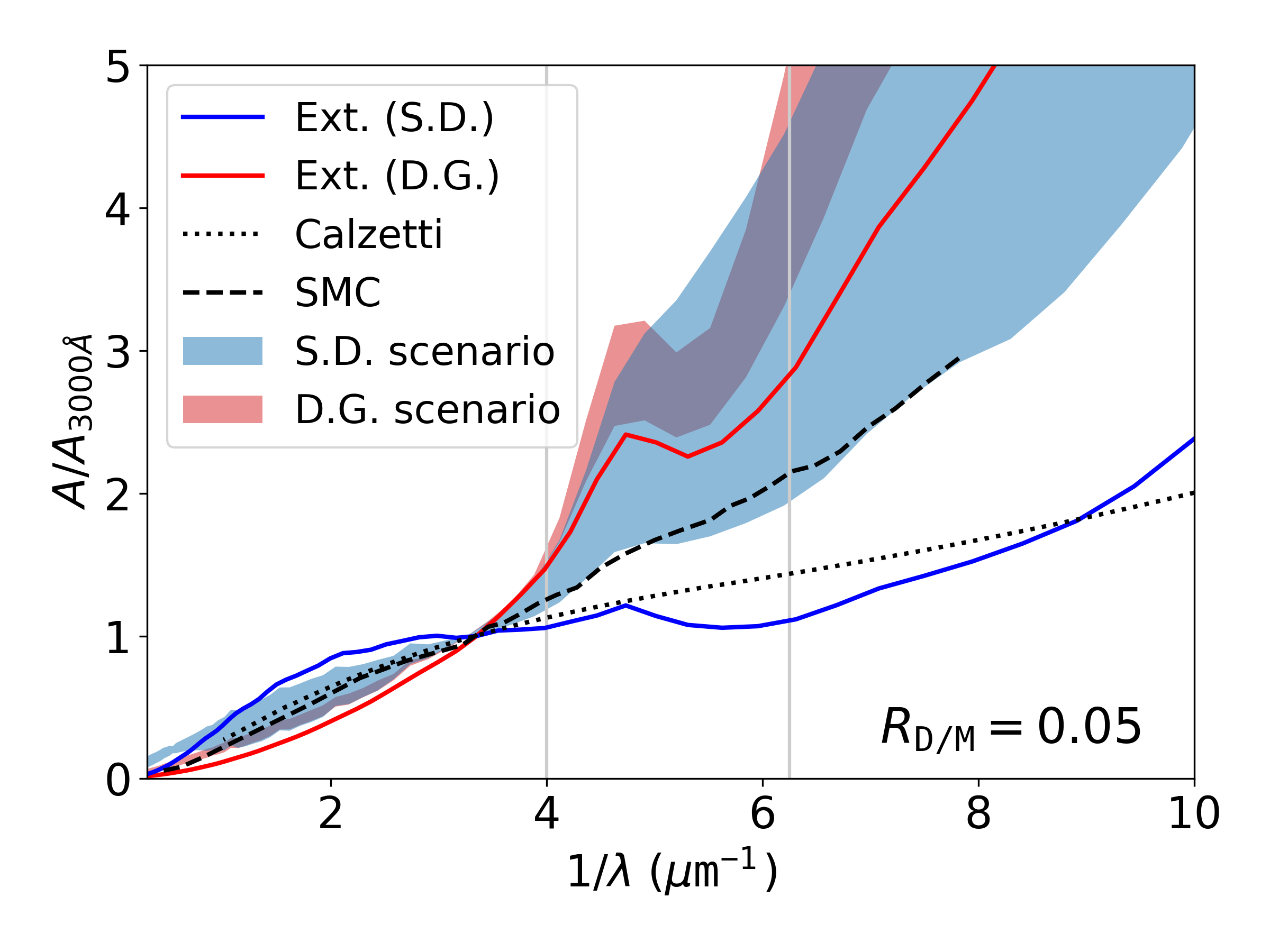}
    \includegraphics[width=\linewidth]{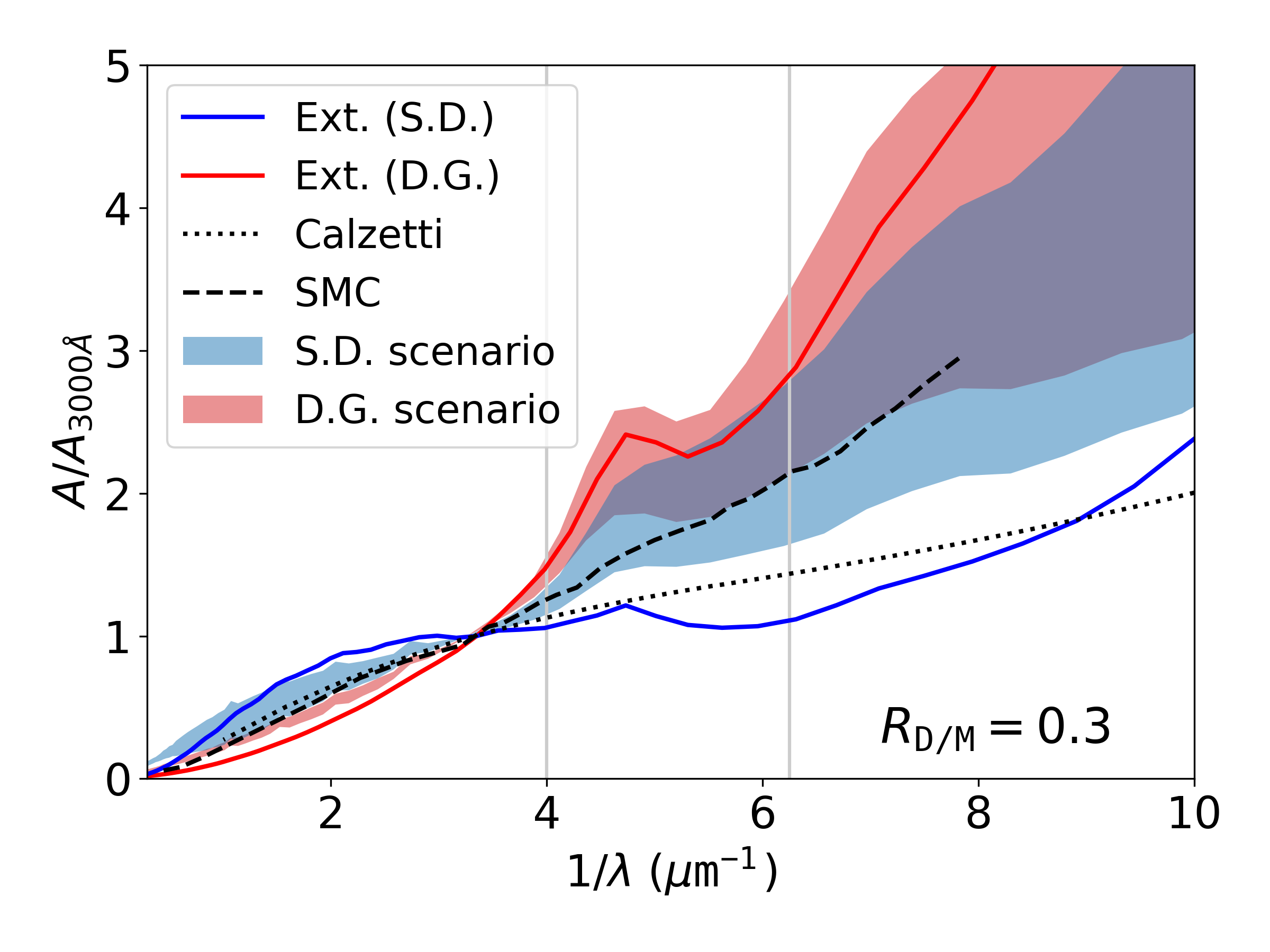}
    \caption{Same as the upper panel of Fig.\ \ref{fig:Att_Cur} but for $R_\text{D/M}=0.05$ (upper) and $0.3$ (lower).}
    \label{fig:Att_dm}
\end{figure}

The upper panel of Fig.\ \ref{fig:Att_Cur} shows the attenuation curves of the TNG sample. The dust growth scenario tends to have steeper attenuation curves than the stardust scenario as expected from the difference in the extinction curves, which are also drawn in the figure. The attenuation curves are similar to the original extinction curve in the dust growth scenario. Therefore, the information on the steep extinction curve is preserved when we observe the attenuation curves. In the stardust scenario, in constrast, the attenuation curves are much steeper than the original extinction curve. The age and scattering effects (see the Introduction and \citetalias{Lin:2021a}) can be responsible for the steepening. We confirmed that the scattering effect is particularly important by `turning off' the scattering in the \textsc{skirt9} calculations. Scattering effectively increases the light path of short-wavelength photons, and raises the probability of them being absorbed by dust.

Although there is a systematic difference in the attenuation curves for the two scenarios, they significantly overlap. Thus, it is practically difficult to distinguish the two scenarios just by comparing the attenuation curves. In both scenarios, the attenuation curves can be even steeper than the SMC extinction curve.

The lower panel of Fig.\ \ref{fig:Att_Cur} shows the IRX--$\beta$ relation of the selected galaxies. We find that the dust growth scenario has higher $\beta$ values and slightly higher IRX values. The change of $\beta$ between the two scenarios is large because the dust growth scenario has a steeper extinction curve, which makes the UV colour redder. Although the attenuation curves heavily overlap as shown in the upper panel, there is a difference in $\beta$ between the two scenarios, since $\beta$ is more sensitive to the extinction curve than IRX.

For comparison, we also plot the expected IRX--$\beta$ relations for two attenuation curves shown in the upper panel: the Calzetti and SMC curves. The IRX--$\beta$ relations for these curves are calculated by attenuating the intrinsic SED $L_\lambda\propto\lambda^{\beta_\mathrm{int}}$ with these attenuation/extinction curves and assuming that the attenuated energy is reradiated in the IR. We adopt a bluer intrinsic value of $\beta (=-2.5)$ compared with that in nearby galaxies ($\simeq -2.2$; \citealt{Meurer:1999a}) to explain the IRX--$\beta$ relation shifted towards smaller $\beta$ \citep{Wilkins:2013a}. Such a blue intrinsic UV colour is also used by \citet{Schouws:2022a} based on a young, low-metallicity stellar population \citep{Reddy:2018a}, and is consistent with comprehensive analysis of UV luminosity functions and UV colours for high-redsift LBGs by \citet{Mancini:2016a}. We find that the data of both scenarios are broadly consistent with the IRX--$\beta$ relation expected from the SMC extinction curve. This reflects the steep attenuation curves in both scenarios. There is a slight trend that the points in the dust growth scenario are located below the dashed line (i.e. systematically larger $\beta$) because the attenuation curves are even steeper than the SMC curve.

For reference, we plot the observed IRX--$\beta$ relations of LBGs at $z\sim 7$ taken from \citet{Hashimoto:2019a}, \citet{Schouws:2022a} and \citet{Bowler:2022a}. Since the errors are large, we only use the observational data for a `guide' to confirm that the theoretical predictions are not deviated significantly from the observed trend. Fig.\ \ref{fig:Att_Cur} shows that the observed data with relatively low IRX are well explained by our models. However, our models are not able to reproduce the data with large IRX. This is slightly improved by higher value of $R_\mathrm{D/M}$, which we discuss in Section \ref{subsec:dm}.

\subsection{Effect of the dust-to-metal ratio}\label{subsec:dm}

We examine the effect of different dust-to-metal ratios ($R_\mathrm{D/M})$. Fig.~\ref{fig:Att_dm} compares the attenuation curves for $R_\mathrm{D/M}=0.05$ and 0.3, smaller and larger values than that shown in Fig.\ \ref{fig:Att_Cur}, respectively. The difference between the two scenarios is qualitatively similar in all cases for $R_\mathrm{R/M}$; that is, the dust growth scenario tends to predict steeper attenuation curves, but the attenuation curves of the two scenarios heavily overlap. However, the range significantly varies depending on $R_\mathrm{D/M}$. As $R_\mathrm{D/M}$ becomes high, the attenuation curves tend to be flat. The flattening of attenuation curves with increasing optical depth (or dust abundance) is consistent with the tendency expected from a well mixed distribution of stars and dust: In this geometry, a certain fraction of stellar UV light easily escape from the galaxy (geometry effect; see the Introduction). We note that the same effect (flattening) is also expected if the stellar and dust emissions are displaced \citep[e.g.][]{Narayanan:2018a,Pallottini:2022a}. However, such a displacement is not clear for our sample after the inspection of individual images of star and dust emissions (but we could not address a displacement below the spatial resolution; Section \ref{subsec:resolution}). Thus, the balance between the flattening and steepening effects regulates the variety in the attenuation curve shapes, with the steepening/flattening prominent in the low/high dust abundance (or optical depth).

\begin{figure}
    \centering
    \includegraphics[width=\linewidth]{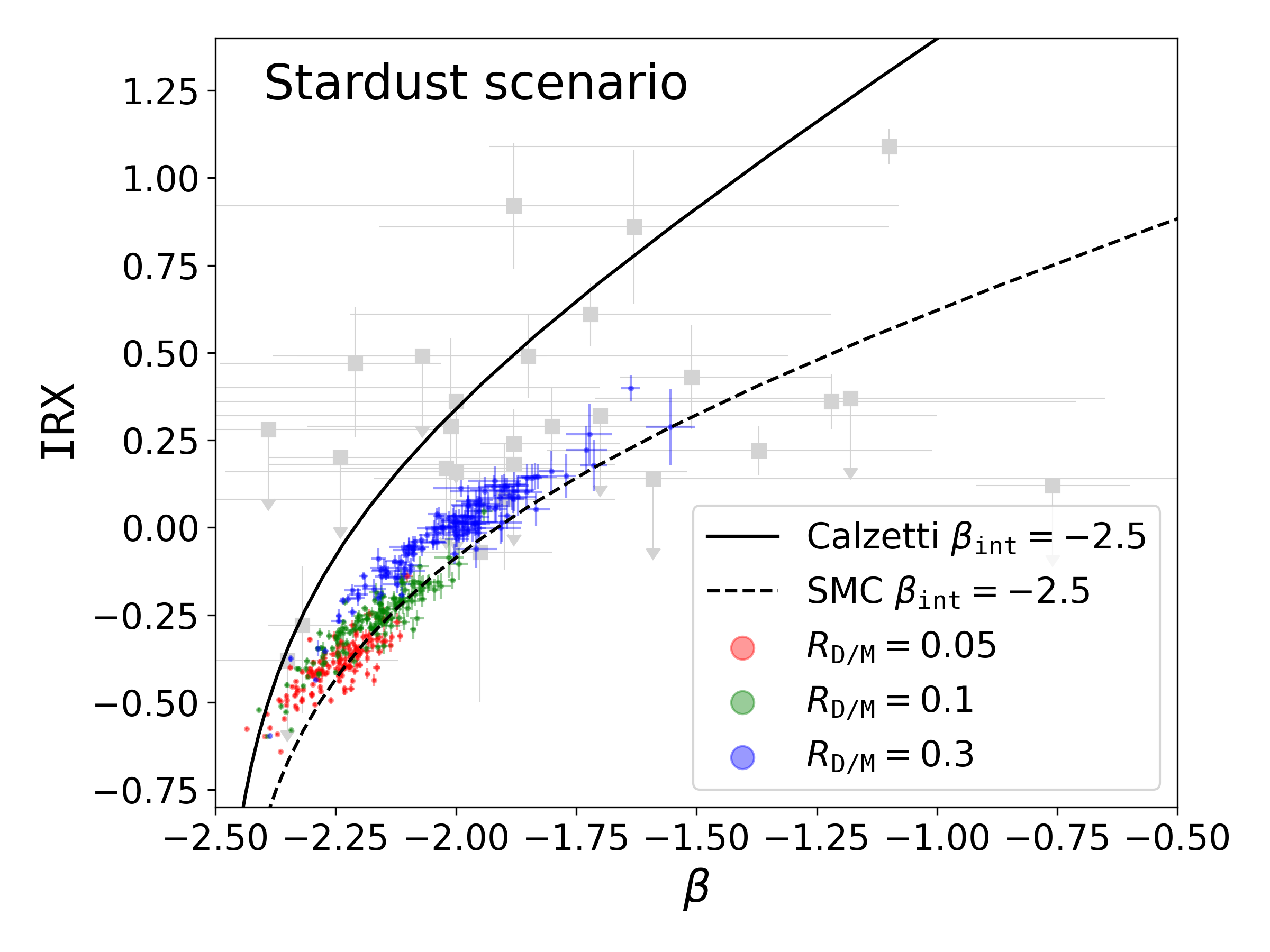}
    \includegraphics[width=\linewidth]{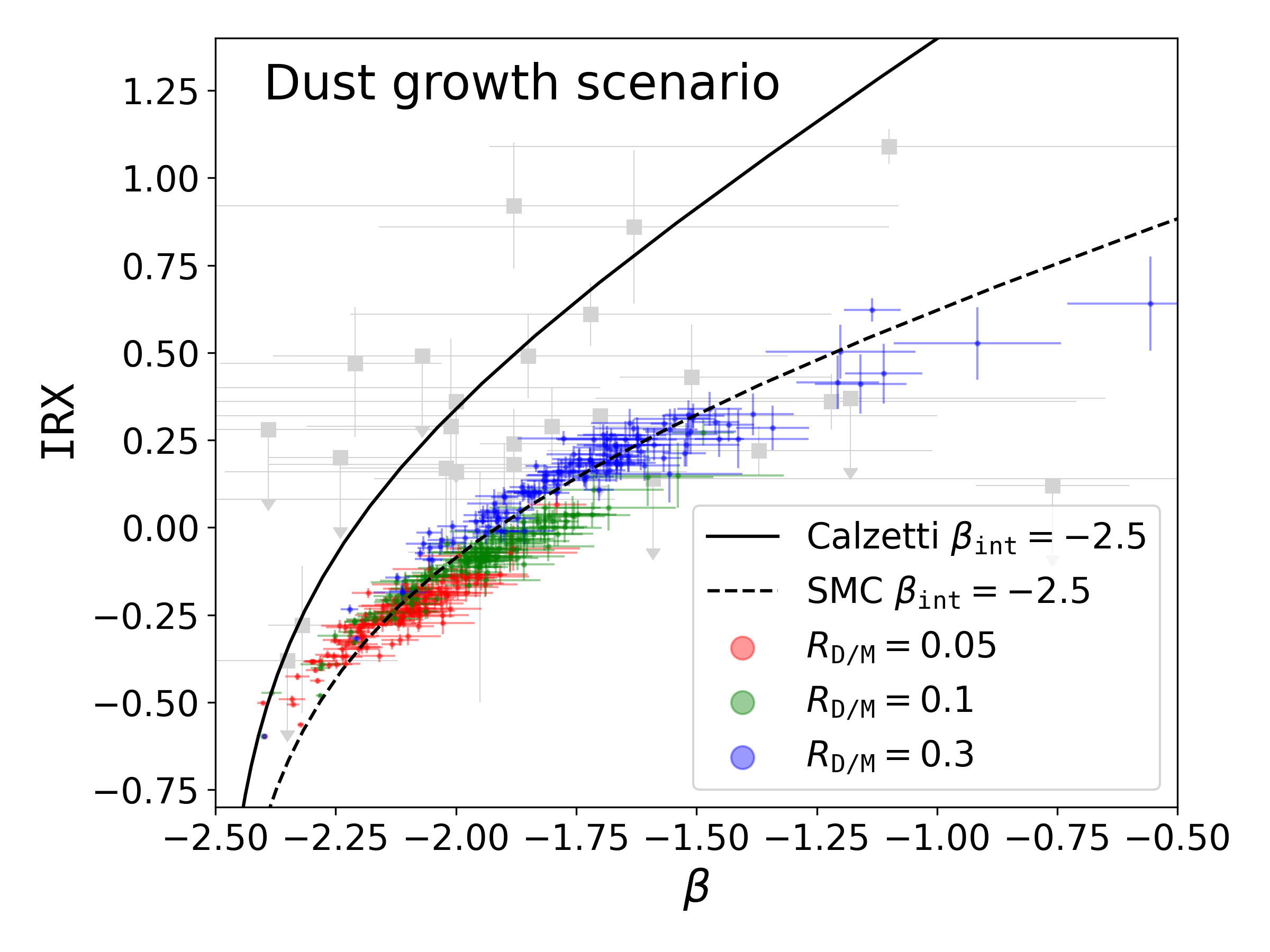}
    \caption{Same as the lower panel of Fig.\ \ref{fig:Att_Cur} but with different dust-to-metal ratios. The upper and lower panels are for the stardust and dust growth scenarios, respectively. The red, green and blue data points represent $R_\mathrm{D/M}=0.05$, $0.1$ and $0.3$ respectively.}
    \label{fig:IRXB_dm}
\end{figure}

In Fig.\ \ref{fig:IRXB_dm}, we show the IRX--$\beta$ relations for $R_\mathrm{D/M}=0.05$, 0.1 and 0.3 in the two scenarios separately. As expected, higher dust-to-metal ratios lead to larger IRX and $\beta$ values. In each scenario, the IRX--$\beta$ relations with different $R_\mathrm{D/M}$ are along a single sequence near to that predicted from the SMC extinction curve. This is partly in line with the conclusion from \citet{Ma:2019a}, who argued that the IRX--$\beta$ sequence is not sensitive to the dust-to-metal ratio. However, the location along the sequence systematically depends on the value of $R_\mathrm{D/M}$.

We also note that the slope of the IRX--$\beta$ relation is different between the two scenarios: In the stardust scenario, the points tend to be located above the SMC line (dashed line), while in the dust growth scenario, they are rather located below. Therefore, a large $R_\mathrm{D/M}$ in the stardust scenario could explain the data points with small $\beta$ and large IRX, while a large $R_\mathrm{D/M}$ in the dust growth scenario is more suitable to explain those with large $\beta$ with small IRX. Therefore, the difference in the original extinction curve (or the dust source) has some impact on the location in the IRX--$\beta$ relation.

In Fig.\ \ref{fig:IRXB_dm}, the same observational data as in Fig.\ \ref{fig:Att_Cur} (lower) are also plotted. Although the two scenarios with different $R_\mathrm{D/M}$ cover a large region consistent with some observational data in the IRX--$\beta$ diagram, all the dispersion in the observational data, especially at low $\beta$ and large IRX, is difficult to explain with our models. This discrepancy could partly be due to the large errors in the observational data, and partly be due to the uncertainties in the theoretical model. As shown in Appendix \ref{apdx:sgratio}, a smaller silicate fraction could increase IRX and decrease $\beta$, although it is still difficult to boost IRX drastically. Another possibility is a mismatch in the stellar mass: Our sample has only a small number of galaxies at $\log (M_\star /\mathrm{M}_{\sun})\sim 9$--10, which is typical for ALMA-detected LBGs. In fact, as discussed in Section \ref{subsec:Mstar}, such massive LBGs may have systematically high dust optical depths, so that high IRX values could be achieved. Some studies focusing on massive objects (e.g.\ by using zoom-in cosmological simulations) could help to resolve this issue.

At slightly lower redshifts ($z\sim 5$), however, some observational data put a strong constraint on the IRX--$\beta$ relation. Stacked data in \citet{Fudamoto:2020a} show the IRX ($\lesssim 0.0$) of $z\sim 5$ LBGs is low even if some of them have $\beta\sim -1.5$. Seeing Fig.\ \ref{fig:IRXB_dm}, such a combination of low IRX and large $\beta$ can be more easily explained by the dust growth scenario. This implies that the production of small grains is already efficient in those LBGs.

A larger contribution from an old stellar population would lead to larger $\beta$ \citep{Narayanan:2018b,Burgarella:2022a}. This could give an alternative explanation for objects with large $\beta$ and small IRX. Thus, the star formation histories, which depend on the adopted star formation and stellar feedback recipe, should also be carefully examined and tested against future observations by e.g.\ \textit{James Webb Space Telescope}.

\subsection{Stellar mass dependence}\label{subsec:Mstar}

\begin{figure}
    \centering
    \includegraphics[width=\linewidth]{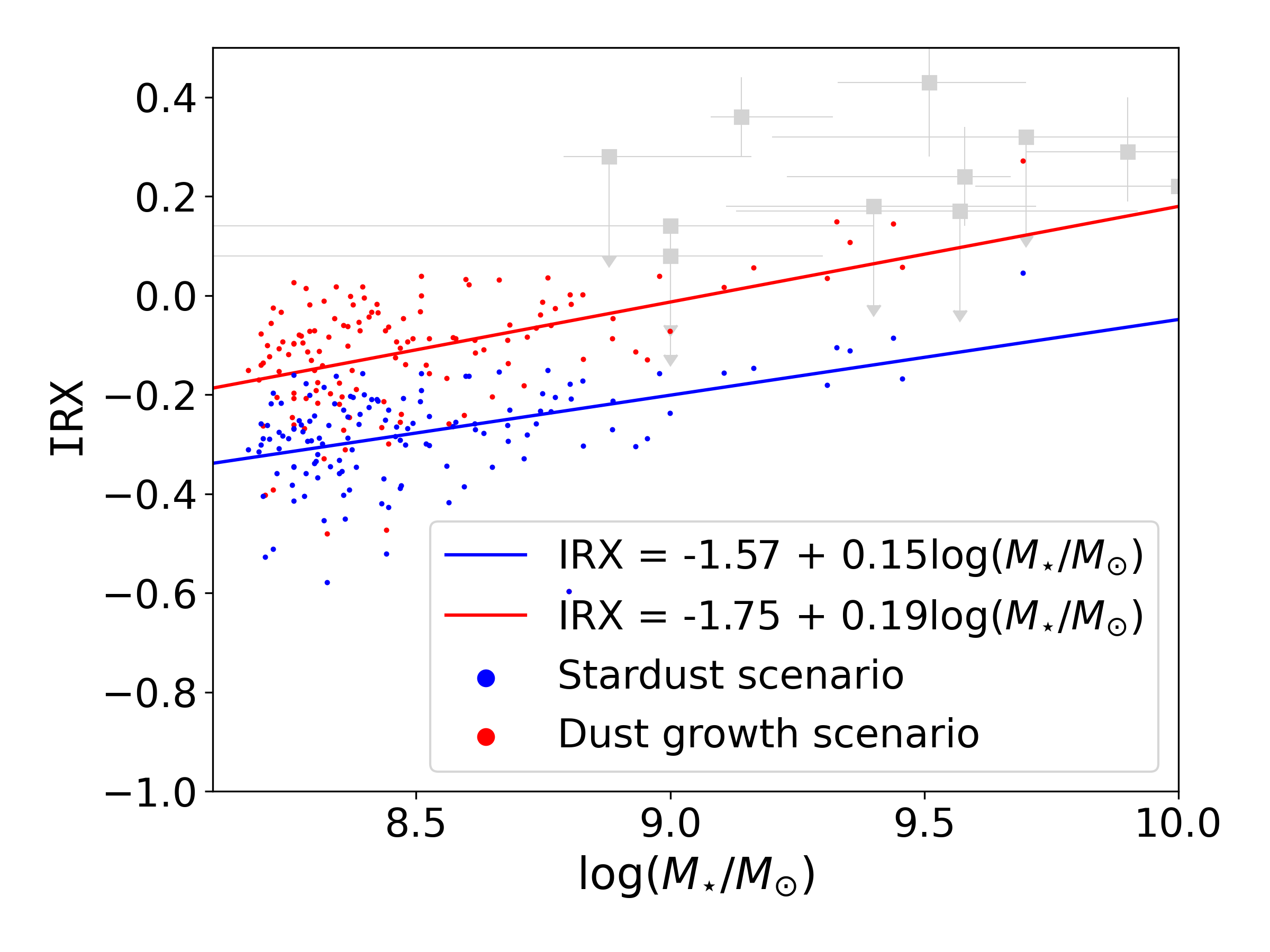}
    \caption{Relations between IRX and stellar mass with $R_\mathrm{D/M}=0.1$. The blue and red points show the results for the stardust and dust growth scenarios, respectively. The linear regression for each scenario is shown by the line in the same colour. The grey points with error bars or arrows (for upper limits) represent the observational data from \citet{Schouws:2022a}. }
    \label{fig:IRX_Mstr}
\end{figure}

In order to examine possible bias caused by the stellar mass selection, we show the stellar mass dependence of IRX. Indeed, at $z\sim 5$, \citet{Fudamoto:2020a} showed that IRX depends on the stellar mass. To address this dependence, we present the relations between IRX and $M_\star$ in Fig.\ \ref{fig:IRX_Mstr}. We adopt $R_\mathrm{D/M}=0.1$. For $R_\mathrm{D/M}=0.05$ (0.3), IRX roughly shifts downwards by 0.1 (upwards by 0.2).

We predict in Fig.\ \ref{fig:IRX_Mstr} a positive correlation between IRX and $M_\star$, although it is difficult to confirm this using the existing data at $z\gtrsim 5$. Such a positive correlation is found at lower redshifts \citep{Heinis:2014a,Bouwens:2016a,Alvarez:2016a,Fudamoto:2020a}. We also observe a small upward offset in the dust growth scenario relative to the stardust scenario. The linear regression is given by $\mathrm{IRX}=-1.57+0.15\log (M_\star/\mathrm{M}_{\sun})$ for the stardust scenario and $\mathrm{IRX}=-1.75+0.19\log (M_\star/\mathrm{M}_{\sun})$ for the dust growth scenario. From these relations, we find the following two points: (i) The dust growth scenario has slightly higher IRX (by $\sim$0.2) since short-wavelength UV light is absorbed more efficiently. (ii) The IRX systematically increases with $M_\star$, and the slope is not significantly different between the two scenarios.

At $z\gtrsim 7$, observational constraints on IRX are difficult to obtain at $M_\star\lesssim 10^9$ M$_{\sun}$, while the small simulation box in TNG50 makes it difficult to sample galaxies at $M_\star\gtrsim 10^{10}$ M$_{\sun}$. Thus, we concentrate on the comparison at $\log (M_\star /\mathrm{M}_{\sun})\sim 9$--10. The resulting IRX values are compared with the observational data of LBGs at $z\sim 7$--8 taken from \citet{Schouws:2022a}. Fig.\ \ref{fig:IRX_Mstr} shows that most of the galaxies at $M_\star\gtrsim 10^9$ M$_{\sun}$ have $\mathrm{IRX}\lesssim 0.2$, which is broadly consistent with the data, especially those with upper limits. Objects with relatively large IRX ($\gtrsim 0.2$) can only be explained by the dust growth scenario. However, our small sample size in the above stellar mass range hampers more detailed comparison. As mentioned above, zoom-in cosmological simulations focusing on massive [$\log (M_\star /\mathrm{M}_{\sun})\sim 9$--10] objects would help to supplement our comparison.

We should keep in mind that the uncertainties in dust temperature affect the estimate of IRX \citep[e.g.][]{Faisst:2017a}. If the dust temperatures are systematically underestimated, the IRX values derived from observed ALMA data are also underestimated. We also note that theoretical studies for IRX have not converged yet. At a fixed $M_\star$, our IRX values are comparable to \citet{Ma:2019a} but are smaller than the results in \citet{Vijayan:2022a}. \citet{Pallottini:2022a} also estimated IRX$\sim 0$--1, systematically larger than our results. Nevertheless, it is difficult to identify the actual reason for the difference because of various underlying assumptions on sub-grid physics \citep[such as feedback models as discussed by][]{Pallottini:2022a}. In the future, we need to investigate the effect of feedback treatments, sub-grid models, and mass resolutions on the resulting IRX using a single simulation framework in order to understand the reason for the discrepancies among the above simulations.

\section{Discussion}\label{sec:discussion}

\subsection{Distinguishing dust sources}

The above results show that the attenuation curves are strongly modified by the radiative transfer effects as already shown by simple dust--stars geometries in  \citetalias{Lin:2021a}. With realistic geometries realized as a result of hydrodynamic evolution, the attenuation curves in the two scenarios (dust sources) are strongly overlapped, so that it is practically impossible to distinguish the dust sources only from the shape of attenuation curve.

The situation is improved if we include IR dust emission in the analysis. The different dust sources (or different extinction curves) cover different areas in the IRX--$\beta$ diagram. In particular, $\beta$ is higher in the dust growth scenario because short-wavelength photons are more efficiently absorbed. Under a certain value of IRX, data points with large (small) $\beta$ are more easily explained by the dust growth (stardust) scenario. Thus, we argue that the statistical distribution of data points in the IRX--$\beta$ diagram reflects the dust source. \citet{Ferrara:2022a} developed an analytic method of deriving various quaitities (e.g.\ dust mass) using IRX and $\beta$, and mentioned that the results depend on the adopted extinction curve. Thus, there is a possibility of constraining the underlying extinction properties of high-redshift galaxies using the IRX--$\beta$ relation.

Although the different dust sources more affect $\beta$, IRX still varies. In particular, using the IRX--$M_\star$ correlation shown in Fig.\ \ref{fig:IRX_Mstr}, we may be able to distinguish the two scenarios statistically. In spite of the fact that the attenuation curves are strongly modified by radiative transfer effects, more detailed analysis using IRX, $\beta$, and galaxy properties (e.g.\ $M_\star$) would lead to a statistical understanding of underlying dust extinction properties, which are affected by the dominant dust sources.

\subsection{Resolution effects}\label{subsec:resolution}

In cosmological simulations, it is difficult to realize small-scale structures such as molecular clouds. As mentioned in Section \ref{subsec:TNG}, our statistical results are not significantly affected by the choice of TNG50 or TNG100, which implies that the resolution in the cosmological simulation is not a critical issue for this paper. Moreover, we have confirmed the convergence in the radiative transfer calculations around the maximum spatial resolution we adopted (a few pc). Therefore, there is no indication that the spatial resolutions strongly affect our conclusions drawn in this paper.

Nevertheless, some small-scale effects that are not dealt with in our treatment could be important. Dust embedded in a dense cloud could be strongly shielded, and such a cloud has a large radial gradient of dust temperature, especially when it is located nearby actively star-forming regions or it hosts intense star formation \citep{Ferrara:2017a,Sommovigo:2020a}. Although we treat such an embedded portion of dust by the MAPPINGS III model (Section \ref{subsec:RT}), the dust temperature still depends on the detailed treatment of small-scale density structures. Dust emission SEDs could be overwhelmed by high-temperature dust, making low-temperature dust `invisible'. Moreover, if small-scale clumpy structures exist, dust and stars could be displaced. Such an displacement could flatten the attenuation curve \citep{Pallottini:2022a}. Thus, the energy balance between dust absorption and dust emission becomes strongly dependent on the small-scale structures. It would thus be useful to combine a large-scale cosmological simulations with a treatment of small-scale structures in future theoretical modelling.

\section{Conclusions}\label{sec:conclusion}

In this work, we investigate the effect of dust sources on attenuation properties in $z\sim7$ galaxies using the TNG50 cosmological simulation data in the IllustrisTNG project. We consider two kinds of grain size distributions, both of which equally explain the total dust abundance of ALMA-detected LBGs at $z\sim 7$ \citep{Liu:2019a}: one is the stardust scenario, in which stellar dust production is the main source of dust, and the other is the dust growth scenario, in which the dominant portion of the dust is formed through the accretion of gas-phase metals in the dense ISM. These two scenarios predict totally different grain size distributions: large ($\gtrsim 0.1~\micron$) and small ($\lesssim 0.01~\micron$) grains are dominant in the stardust and dust growth scenarios, respectively. The extinction curve is flat (steep) in the stardust (dust growth) scenario. We perform radiative transfer calculation with \textsc{skirt9} based on the dust properties in the two scenarios by treating the dust-to-metal ratio, $R_\mathrm{D/M}$, as a free parameter. We output attenuation curves and IRX--$\beta$ relations to examine whether or not dust attenuation properties are different between those two scenarios in the dust and star distributions (geometries) realized as a result of cosmological hydrodynamic simulations.

We confirm that the attenuation curves are substantially different from the original extinction curves because of various radiative transfer effects, which were already clarified using simple spherical geometries by \citetalias{Lin:2021a}. In the stardust scenario, in particular, the attenuation curves are drastically steepened by the effects of scattering and stellar-age-dependent extinction. These effects make the attenuation curves in the two scenarios statistically similar, making it difficult to distinguish the dominant dust sources only from the attenuation curve shapes.

We also examine the IRX--$\beta$ relation, which is often used to clarify the attenuation properties of galaxies. A large number of the data points are located near to the IRX--$\beta$ relation expected from the SMC extinction curve with a bluer instinsic stellar UV colour than is adopted for nearby galaxies. We still find a systematic difference in the IRX--$\beta$ relation between the two scenarios. The dust growth scenario has systematically higher $\beta$ (i.e.\ redder UV colours) than the stardust scenario. In particular, the objects with $\beta >-2.0$ with $\mathrm{IRX}\lesssim 0$ are more easily explained by the dust growth scenario. This is because the attenuation curves in the dust growth scenario is steeper at a fixed value of IRX.

We also vary the dust abundance by changing the dust-to-metal ratio $R_\mathrm{D/M}$. There is a trend of steeper attenuation curves for smaller $R_\mathrm{D/M}$. This is because the steepening effects (mainly due to scattering, which increases the chance of absorption particularly at short wavelengths) appear more easily at low optical depth, while the flattening effect by well mixed dust--stars geometries becomes more prominent at high optical depth. The IRX--$\beta$ relation also shifts towards smaller $\beta$ and smaller IRX for smaller $R_\mathrm{D/M}$ as expected from the change of dust optical depth. As mentioned above, the dust growth scenario tends to show larger $\beta$ at a fixed IRX than the stardust scenario in all values of $R_\mathrm{D/M}$. Thus, the wide variety in the observed IRX--$\beta$ relation at $z\sim 7$ could be explained to some extent by different dust formation scenarios together with a variety in $R_\mathrm{D/M}$: Objects with blue (red) UV colours but high (low) IRX values are more easily explained by the stardust (dust growth) scenario. It is, however, difficult to explain the observational data points with $\mathrm{IRX}\gtrsim 0.5$ with our models.

The IRX--$M_\star$ relation is also influenced by the dust sources (extinction curves). The dust growth scenario has a significantly higher IRX at a fixed $M_\star$ than the stardust scenario. This indicates that the relation between IRX (or $\beta$) and galaxy properties (e.g.\ $M_\star$) as well as the IRX--$\beta$ relation gives a useful clue to statistically distinguishing different extinction curves (dust sources).

We conclude that the dust--stars geometry in a galaxy actively shapes the attenuation curve. The attenuation curves tend to become similar even if the original extinction curves are very different. This effect makes it difficult to distinguish the dominant dust sources only with the attenuation curve shape. However, the two scenarios (with different dust sources) still cover different regions on the IRX--$\beta$ diagram. Thus, the IRX--$\beta$ relation is still useful in constraining the dust properties predicted from various dust enrichment models.

\section*{Acknowledgements}
We thank Y.-H. Huang for helping us with numerical setups, D. Nelson and A. Pillepich for their advice on the TNG50 data, and the anonymous referee for useful comments. We appreciate the IllustrisTNG collaboration for providing free access to the data used in this paper.
Numerical computations were carried out on XL at the Theoretical Institute for Advanced Research in Astrophysics (TIARA) in Academia Sinica.
HH thanks the Ministry of Science and Technology for support through grant 108-2112-M-001-007-MY3, the National Science and Technology Council for grant 111-2112-M-001-038-MY3, and the Academia Sinica for Investigator Award AS-IA-109-M02.

\section*{Data Availability}
Data related to this publication and its figures are available on request from the corresponding author. The IllustrisTNG simulations are publicly available and accessible at \url{www.tng-project.org/data} \citep{Nelson:2019a}.



\bibliographystyle{mnras}
\bibliography{references} 




\appendix

\section{Influence of the silicate fraction} \label{apdx:sgratio}
We examine the influence of dust composition by changing the silicate dust mass fraction $f_\mathrm{sil}$. Besides a run with $f_\mathrm{sil}=0.9$ discussed in the text, we ran calculations with $f_\mathrm{sil}=0.6$ and 0.8 to cover the range of the silicate fraction expected for various galaxy ages \citep{Hirashita:2020a}.

\begin{figure}
    \centering
    \includegraphics[width=\linewidth]{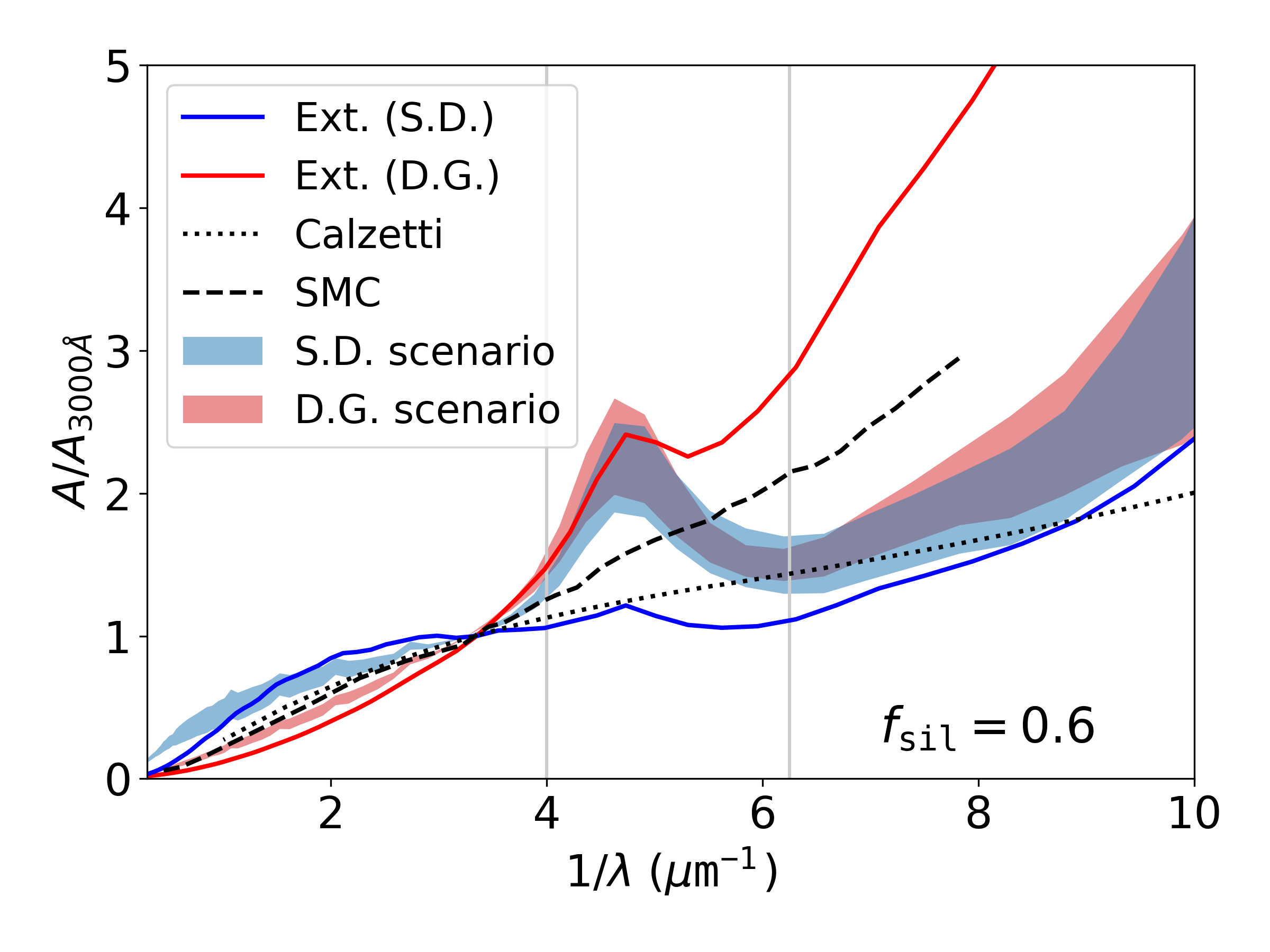}
    \caption{Same as the upper panel of Fig.\ \ref{fig:Att_Cur} but for $f_\mathrm{sil}=0.6$. The blue and red solid curves show the original extinction curves with $f_\mathrm{sil}=0.6$ for the stardust and dust growth scenarios, respectively.}
    \label{fig:Att_SG}
\end{figure}

We show the attenuation curves with $f_\mathrm{sil}=0.6$ in Fig.\ \ref{fig:Att_SG}. Both of the extinction and attenuation curves are flattened compared with those shown in Fig.\ \ref{fig:Att_Cur} because of less contribution from silicate, which produced steeply rising far-UV extinction curve. We still observe that the attenuation curve shapes are heavily overlapped between the two scenarios because the attenuation curve is steeper (flatter) than the extinction curve in the stardust (dust growth) scenario.

\begin{figure}
    \centering
    \includegraphics[width=\linewidth]{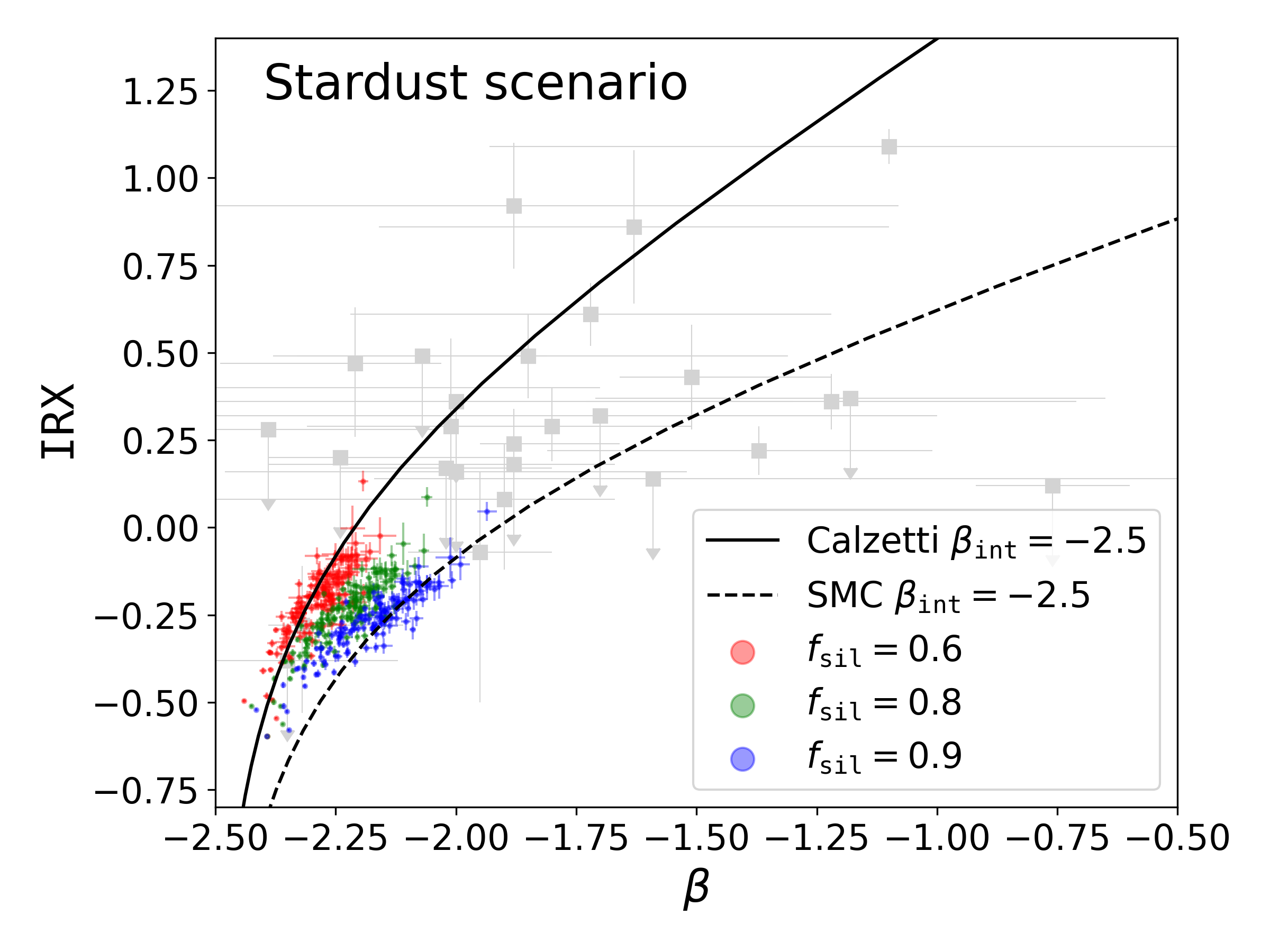}
    \includegraphics[width=\linewidth]{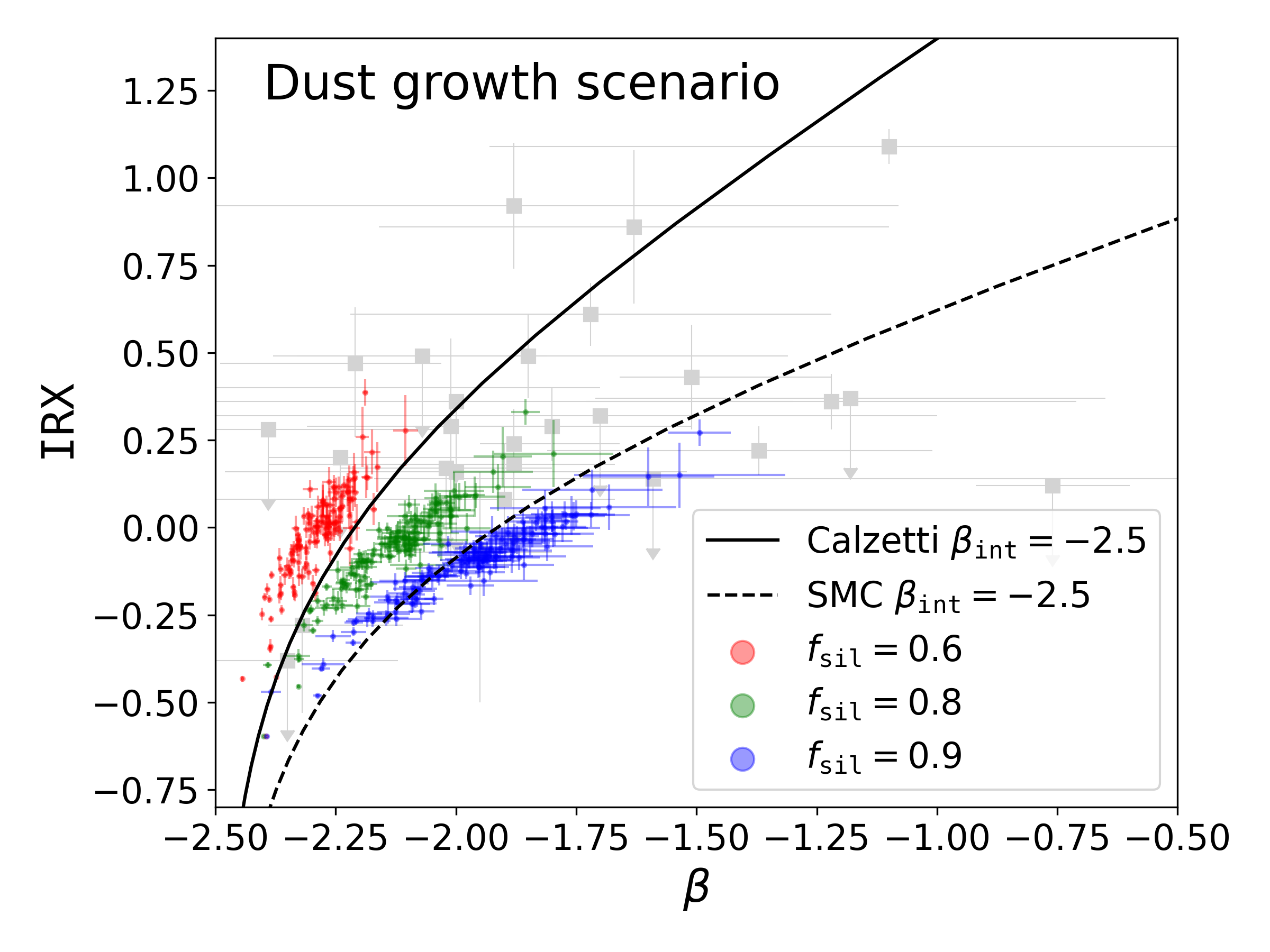}
    \caption{Same as the lower panel of Fig.\ \ref{fig:Att_Cur} but with different values for the silicate dust mass fraction ($f_\mathrm{sil}$). The upper and lower panels are for the stardust and dust growth scenarios, respectively. The red, green and blue data points represent $f_\mathrm{sil}=0.6$, $0.8$ and $0.9$ respectively.}
    \label{fig:IRXB_SG}
\end{figure}

We show the IRX--$\beta$ relations adopting these values of  $f_\mathrm{sil}$ for the two scenarios in Fig.\ \ref{fig:IRXB_SG}. Since we only adopted $0.16~\micron$ and $0.25~\micron$ for calculate the IRX--$\beta$ relations, the strong carbon bump at $0.22~\micron$ in the attenuation curves does not directly affect our discussion. We observe that the change of $f_\mathrm{sil}$ moves the IRX--$\beta$ relation to the direction roughly vertical to the well identified IRX--$\beta$ sequence. For a fixed value of $f_\mathrm{sil}$, the IRX--$\beta$ relation still extends more towards large $\beta$ in the dust growth scenario than in the stardust scenario. The variation of $f_\mathrm{sil}$ contributes to explaining the scatter towards larger values of IRX at a certain value of $\beta$. The IRX--$\beta$ relation for $f_\mathrm{sil}=0.6$ approaches the sequence predicted with the Calzetti attenuation curve, which reflects the flatter attenuation curves for larger $f_\mathrm{sil}$.

\bsp	
\label{lastpage}
\end{document}